\documentclass[12pt,preprint]{aastex}
\usepackage{emulateapj5}

\shorttitle{SIGMA Observations of GRO J1744-28}
\shortauthors{Mej\'{\i}a et al.}

\input psfig.sty

\begin{document}

\title{SIGMA Observations of the Bursting Pulsar GRO J1744-28}
\author{J. Mej\'{\i}a and T. Villela}
\affil{Instituto Nacional de Pesquisas Espaciais, Divis\~ao de Astrof\'{\i}sica,
C.P. 515, 12201-970, S\~ao Jos\'e dos Campos, SP, Brazil}
\email{mejia@das.inpe.br}

\author{P. Goldoni and F. Lebrun}
\affil{CEA/DSM/DAPNIA/SAp, CEA-Saclay, F-91191 Gif-sur-Yvette Cedex, France}

\author{L. Bouchet, E. Jourdain, J.-P. Roques and P. Mandrou}
\affil{Centre d'Etude Spatiale des Rayonnements, 9 Avenue du Colonel Roche, B.P. 4346, 31029 
Toulouse Cedex, France}

\and

\author{S. Kuznetsov, N. Khavenson, A. Dyachkov, I. Chulkov, B. Novikov, K. Shuhanov, 
I. Tserenin and A. Sheikhet}
\affil{Space Research Institute, Profsoyuznaya 84/32, Moscow 117810, Russia}

\begin{abstract}
We present the results of the GRANAT/SIGMA hard X-/soft $\gamma$-ray 
long-term monitoring of the Galactic Center (GC) region concerning the source 
GRO J1744-28, discovered on 1995 Dec. 2 by CGRO/BATSE. SIGMA observed 
the region containing the source in 14 opportunities 
between 1990 and 1997. In two of these observing sessions, corresponding 
to March 1996 and March 1997, GRO J1744-28 was 
detected with a confidence level greater than 5$\sigma$ in the 35-75 keV 
energy band without detection in the 75-150 keV energy band. For the other 
sessions, upper limits of the flux are indicated. The particular 
imaging capabilities of the SIGMA telescope allow us to identify, 
specifically, the source position in the very crowded GC region, giving 
us a mean flux of (73.1$\pm$5.5)$\times$10$^{-11}$ and (44.7$\pm$6.4)$\times$10$^{-11}$
ergs cm$^{-2}$ s$^{-1}$ in the 35-75 keV energy band, for
the March 1996 and March 1997 observing sessions, respectively. 
Combining the March 1997 SIGMA and BATSE observations, we found evidence pointing to 
the type-II nature of the source bursts for this period. 
For the same observing campaigns, spectra 
were obtained in the 35 to 150 keV energy band. The best fit 
corresponds to an optically thin thermal Bremsstrahlung with 
F$_{50 keV}$=3.6($\pm$0.6)$\times$10$^{-4}$ phot cm$^{-2}$ s$^{-1}$ keV$^{-1}$ 
and kT$_{Bremss}$=28$\pm$7 keV, for the first campaign, and 
F$_{50 keV}$=2.3($\pm$0.7)$\times$10$^{-4}$ phot cm$^{-2}$ 
s$^{-1}$ keV$^{-1}$ and kT$_{Bremss}$=18$^{+12}_{-7}$ keV, for the second. 
This kind of soft spectrum is typical of binary sources containing a neutron 
star as the compact object, in contrast to the harder spectra typical of systems 
containing a black hole candidate. 

\end{abstract}


\section{INTRODUCTION}

GRO J1744-28, the bursting pulsar, was discovered on 1995 December 2 by 
CGRO/ BATSE \citep{fis95,kou96}. Timing studies revealed a period of 
0.467 s in the persistent flux and an 11.8-day orbital period \citep{fin96}. 
This source presented two strong outbursts: 
at the end of 1995 and early 1996, and then almost a year later. At that time, 
this was the only source known to show, simultaneously, periodic 
X-ray pulsations and frequent X-ray bursts. The presence of pulsations 
indicates that the compact object is a neutron star.  The X-ray mass 
function, determined by \citet{fin96} as being f(M$_x$)=1.31$\times$10$^{-4}$ 
M$_{\odot}$, together with a mass canonical value for the compact object of 
1.4 M$_{\odot}$, suggest that the neutron star in GRO J1744-28 is 
being fed by Roche lobe overflow from a low-mass red giant \citep{dau96,mil96,stu96,bil97}.

The best position coordinates for the source are those obtained from 
triangulation of the Ulysses and BATSE observations of the region. With 
this method, \citet{hur00} determined the coordinates RA=17$^h$44$^m$32$^s$, 
decl.=-28$^{\circ}$44$^{'}$31$^{''}.$7 (J2000.0). A possible infrared counterpart was 
reported by \citet{aug96} while a posterior optical/IR 
observation of the same region by \citet{col97} neither 
confirmed nor discarded this IR counterpart. Based on the ASCA data, the
absorption column, N$_H$, for the position 
of GRO J1744-28 was determined as being (5-6)$\times$10$^{22}$ cm$^{-2}$,
independent of the emission date and phase and, therefore, corresponding to 
interstellar absorption \citep{dot96,nis99}. This suggests 
that the source is located near the Galactic Center (GC), at a distance of $\sim$8.0 kpc.
The source was also observed with GRANAT/WATCH from January 13 to March 12, 1996
\citep{saz97} showing, at the maximum between bursts, a daily average flux 
of $\sim$3.7 Crab, in the 8-20 keV energy band.

Between 1990 and 1997, the SIGMA telescope was pointed in the direction 
of the GC region two times per year, in the March-April and 
September-October periods. In 14 opportunities, the region containing 
GRO J1744-28 was observed, for a total of $\sim$2993 hours. 
In two of these sessions, corresponding to March 1996 and March 1997, 
the source was clearly visible, with a confidence level higher than 5$\sigma$. In 
this paper, the analysis of these observations is presented. Also, 
upper limits are reported for the remaining observing sessions. 
Finally, for the March 1997 observing session, a combination of the 
SIGMA value for the integrated flux of the source with the BATSE value 
for the burst mean flux allows us to obtain evidence favouring 
the type-II nature of the bursts, due to spasmodic accretion rather than 
thermonuclear burning of matter.

\section{THE SIGMA TELESCOPE}

The French coded-mask telescope SIGMA \citep{pau91} onboard the
Russian GRANAT orbital observatory provides high resolution images in the hard
X-/soft $\gamma$-ray band from 35 keV to 1300 keV, divided in 95 energy 
channels. The nominal angular resolution of the instrument is $\sim$15 
arcmin. The position determination accuracy of the instrument is 3-5 
arcmin for a 6$\sigma$ source and $<$1 arcmin for a 30$\sigma$ one. The fully 
coded field of view of the instrument is a 4$^{\circ}$.7 x 4$^{\circ}$.3 
rectangle while the half sensitivity boundary of the field of view corresponds 
to an 11$^{\circ}$.5 x 10$^{\circ}$.9 rectangle. The energy resolution of 
the instrument at 511 keV is $\sim$8$\%$. The typical duration of individual 
observing sessions is $\sim$20 hours, that provides a 1$\sigma$ 
sensitivity of $\sim$20 mCrab in the 35-150 keV energy domain. For details on
the SIGMA in-orbit performances, see \citet{bou01}.

\section{THE OBSERVATIONS}

The region containing the source GRO J1744-28 was 
observed by SIGMA in 14 different opportunities, from 1990 to 1997. In two of these 
observing campaigns, the source was clearly detected in the 35-75 keV energy 
band, with a confidence level higher than 5$\sigma$, as shown in Figure 1. 
In contrast, the source was not detected in the 75-150 keV energy band. 
The first campaign began on 1996, March 15 (MJD 50157.63), 103 days 
after the discovery of the source by BATSE, and lasted until March 30, 
corresponding to $\sim$267 hours of effective observing time (see Table 1).
In this period, GRO J1744-28 appears as a 13.5$\sigma$ source
with a mean detected flux in the 35-75 keV energy band of 85$\pm$6.4 mCrab (7.3($\pm$0.6)$\times$10$^{-10}$ 
ergs cm$^{-2}$ s$^{-1}$), 
showing a soft spectrum that can be approximated by an optically thin 
thermal Bremsstrahlung model ($\alpha$=-1.4) with a flux at 50 keV of F$_{50 keV}$= 
3.6($\pm$0.6)$\times$10$^{-4}$ 
phot cm$^{-2}$ s$^{-1}$ keV$^{-1}$ and kT$_{Bremss}$=28$\pm$7 keV 
($\chi^2$(d.o.f.)=13.0(8)).

The second campaign began on 1997, March 14 (MJD 50521.31) and last 
until March 28, corresponding to an effective observational time of $\sim$148.5 
hours. In this opportunity, the mean flux of the source in the 35-75 keV 
energy band was of 52$\pm$7.4 mCrab (4.5($\pm$0.6)$\times$10$^{-10}$ ergs cm$^{-2}$ s$^{-1}$), 
corresponding to a confidence level of 8$\sigma$. The source showed a very steep 
spectrum fitting with an optically thin thermal Bremsstrahlung 
model with a F$_{50 keV}$=2.3($\pm$0.7)$\times$10$^{-4}$ phot cm$^{-2}$ s$^{-1}$ 
keV$^{-1}$ and kT$_{Bremss}$=18$^{+12}_{-7}$ keV ($\chi^2$(d.o.f.) = 8.04(8)). 
In Table 1, the values of the fluxes 
corresponding to each one of the sessions of these two campaigns in the 35-75 keV 
and 75-150 keV energy bands are indicated.

\section{THE LIGHT CURVE AND SPECTRA OF GRO J1744-28}

In Figure 2, the light curves of the source GRO J1744-28 for the 
campaigns corresponding to March 1996 and March 1997 are presented. In the 
first campaign, the source flux in the 35-75 keV energy band was almost 
constant with an average flux of $\sim$85 mCrab or 7.3$\times$10$^{-10}$
ergs cm$^{-2}$ s$^{-1}$. The second campaign does not show significant flux variations; 
the average flux being $\sim$52 mCrab or 4.5$\times$10$^{-10}$
ergs cm$^{-2}$ s$^{-1}$ (see also \citet{tru99}. This last value 
is a factor 7 lower than that reported by \citet{woo99} of
$\sim$3$\times$10$^{-9}$ ergs cm$^{-2}$ s$^{-1}$, as taken from the 
BATSE observations of the source using the Earth-occultation technique 
in the 30 to 100 keV energy band. 

Since GRO J1744-28 is located near the GC, it was within the SIGMA 
field of view during all the observations constituting the GC survey 
conducted by SIGMA between 1990 and 1997, which allows us to estimate upper 
limits for the hard X-ray flux from the source position before and after 
the 1996 and 1997 outbursts. An analysis of these SIGMA observations 
was done in order to verify the possibility of source appearances in 
other epochs than that corresponding to the two outbursts on 1996 
and 1997. In Figure 3, the light curve of the source is shown, 
corresponding to all the pointed SIGMA observations of the GC region containing 
the source, in the 35-75 keV energy band. The resulting 3$\sigma$
upper limit for the average source flux during each campaign is $\sim$17 mCrab 
(1.5$\times$10$^{-10}$ ergs cm$^{-2}$ s$^{-1}$) varying between 9 and 24 mCrab, 
with an integrated upper limit of $\sim$5 mCrab (4.3$\times$10$^{-11}$ ergs cm$^{-2}$ s$^{-1}$).

As said previously, for the two observing sessions in which GRO J1744-28 
was detected, this source presented a very steep spectrum, reasonably well fitted 
with an optically thin thermal Bremsstrahlung model, as given by the expression

$$\frac{dN}{dE} \propto E^{-1.4} \; exp \left( \frac{-E}{kT} \right).$$ 

In order to have a better coverage of 
the spectral behaviour of the source, we also analyzed the RXTE archival data obtained 
on 1996 March 22 and 1997 March 18. The source was very bright during RXTE/PCA observations,
therefore any possible contamination by the other sources in the PCA's field of view is
negligible. The PCA data reduction was performed with the help of standard tools from
LHEASOFT/FTOOLS 5.0.4 package. Both RXTE/PCA and GRANAT/SIGMA spectra were approximated by
a model which is typical for pulsars: power law with high-energy cutoff, Gaussian line and 
photoabsorption. The pure statistical significance of SIGMA data is much less than that of
RXTE/PCA. Because of this, the SIGMA spectral points almost do not influence on the parameters 
of the combined PCA+SIGMA spectral approximation. In order to make the spectral fit more 
sensitive to the SIGMA data, we artificially increased the uncertainties of the PCA spectral
points to the value 5\%. The best fit parameters of the combined PCA+SIGMA spectral fits are
(for 1996/1997): photon index of the power law $\alpha$=1.44$\pm$0.12/1.41$\pm$0.07; 
E$_{Hcutoff}$=14.7$\pm$0.7/16.4$\pm$0.9. In Figure 4 we plot the 1 to 100 keV PCA/SIGMA 
spectrum corresponding to the March 1996 and March 1997 SIGMA observing periods, along 
with the SIGMA spectral fit.

Our result for the second outburst is compatible with the analysis of the same 
event of the source made on the BATSE data and presented by \citet{woo99}. 
For the epoch of the SIGMA observation, they report a temperature 
kT $\sim$14.5$\pm$1.8 keV for the persistent emission (see Figure 4, therein). 
However, as it will be shown later, there exists the possibility of a contamination 
of the BATSE data because of the presence of a transient source in the neighbourhood of GRO J1744-28,
namely GRS 1737-31.

\section{CONCLUSION}

In the period going from 1990 to 1997, observations of the GC region 
containing the coordinates of GRO J1744-28 were made with the SIGMA 
telescope in 14 opportunities, in March-April and September-October each 
year. In two of these opportunities, corresponding to March 1996 and March 
1997, GRO J1744-28 was clearly detected 
well above the 5$\sigma$ flux limit, corresponding to the two outbursts 
detected by several other instruments.
 
The mean fluxes for these two observations, as seen by SIGMA in the 
35-75 keV energy band, were 85 and 52 mCrab. In the 75-150 keV energy 
band, the source was not detected. The spectra were reasonably well 
fitted by optically thin thermal Bremsstrahlung with kT$_{Bremss}$=28 and 18 
keV respectively. These values are compatible with that presented by \citet{woo99} 
for the second burst. However, the flux seen by SIGMA during this second 
observation campaign is almost a factor 7 smaller than that based on the BATSE
 observations of the region using the 
Earth-occultation technique. In this epoch, the transient source GRS 
1737-31 was discovered with SIGMA \citep{sun97} and observed by 
RXTE and BeppoSAX/WFC \citep{mar97,cui97,hei97}. 
This source was active for a period of at least 1 month beginning 
on middle February, 1997, in the 35-150 keV energy band, as reported by 
\citet{tru99}. We argue that the presence of this source may 
explain, at least partially, the different flux values obtained with BATSE 
and SIGMA for the bursting pulsar. For GRS 1737-31, \citet{tru99} have 
found a flux of $\sim$100 mCrab (1.6$\times$10$^{-9}$ ergs cm$^{-2}$
 s$^{-1}$) in the 35-150 keV which, added to the GRO J1744-28 flux (4.5$\times$10$^{-10}$ 
ergs cm$^{-2}$ s$^{-1}$), accounts for a representative part of
the flux reported from BATSE. It is important to note that the use of 
coded mask imaging techniques, like that used by SIGMA and other 
instruments of this kind, allows us to obtain in an unambiguous way the 
integrated flux and spectrum of the sources even in very crowded fields as in 
the case of the GC region.

With regard to the nature of the bursts, several authors have proposed 
that they are type-II bursts (see, for instance, Lewin, Rutledge and Kommers 1996 for 
the first outburst). In their Figure 6, \citet{woo99} have shown that the 
parameter alpha, the ratio between the persistent flux and the average burst 
flux taken over the time interval since the previous burst, displayed for 
almost all the BATSE observations of the second 
outburst a value greater than 20, which suggests that they are due to 
thermonuclear burning rather than spasmodic accretion (type-I bursts rather 
than type-II). However, if we consider the value of the flux obtained 
from the SIGMA observations as the sum of the persistent and the burst 
flux, and the value of the burst flux from the BATSE data, the value of 
alpha for the period of the observation is of the order of the 10, 
more consistent with the type-II nature of the bursts, as indicated by \citet{lew95}.
Once again, the difference 
could be due to a possible contamination of the BATSE data by GRS 1737-31. 

\acknowledgments We acknowledge the paramount contribution of the SIGMA
Project Group of the CNES Toulouse Space Center to the overall success
of the mission. We thank the staffs of the Lavotchine Space Company, of
the Babakin Space Center, of the Baikonour Space Center, and the Evpatoria 
Ground Station for their unfailing support. This research has made use of data
obtained through the High Energy Astrophysics Research Center Online Service, 
provided by the NASA/Goddard Space Flight Center. J. Mej\'{\i}a was supported
by CAPES and CAPES grant BEX1224/00-0. T. Villela was partially supported by 
CNPq under grant 302266/88-7-FA. The Space Research Institute (IKI)
co-authors would like to acknowledge partial support of this work by RBRF grant
No. 00-15-96649.

\clearpage

\begin{figure}

\centerline{\psfig{figure=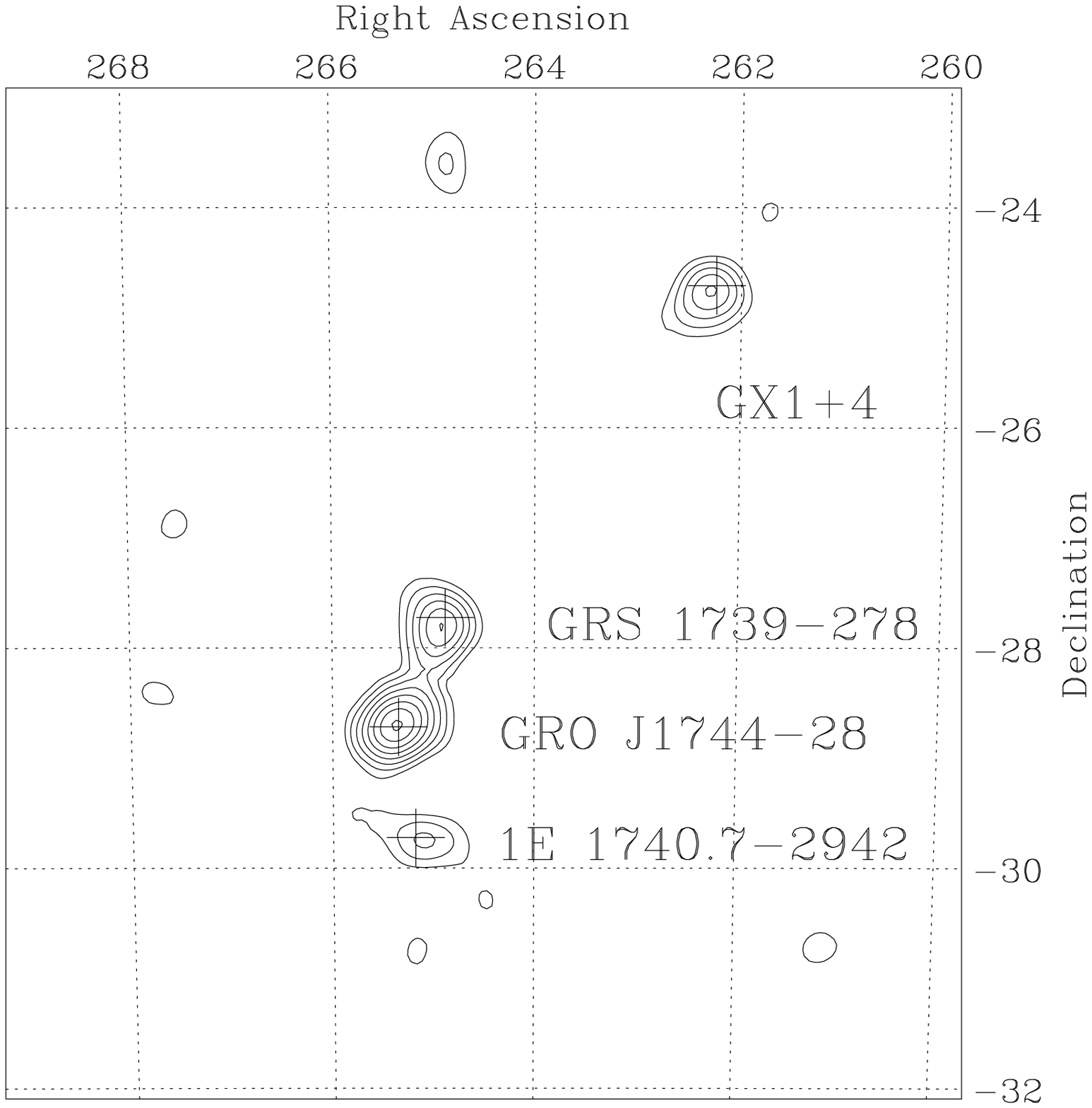,height=75mm,width=75mm}
            \psfig{figure=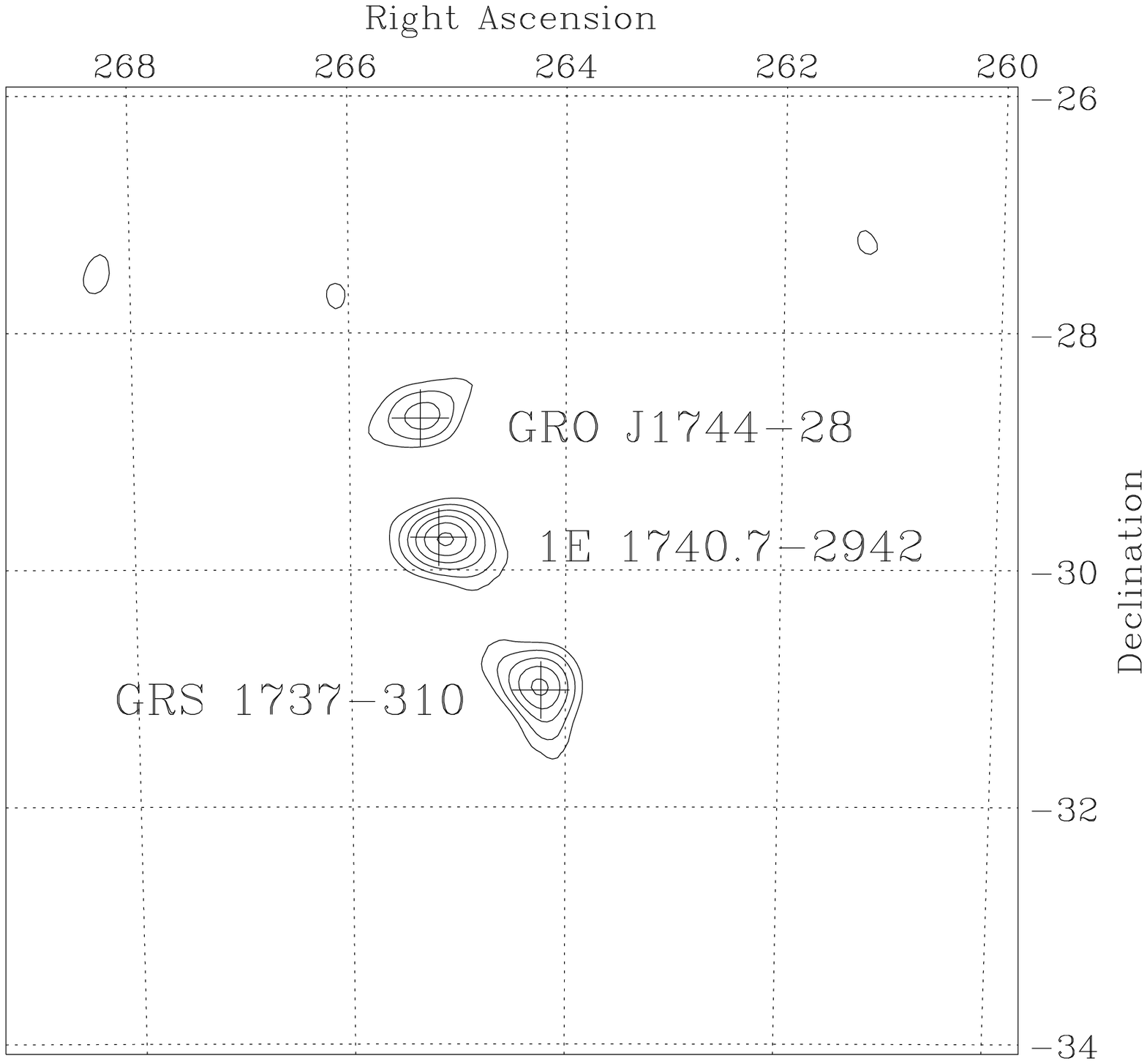,height=75mm,width=75mm,angle=90}}

\caption{Contour images of the GC region obtained with SIGMA on March 1996 (left) and 
March 1997 (right) in the 35-75 keV energy band. Right Ascension and Declination are
in degrees. Notice the presence of the transient 
source GRS 1737-31 in the March 1997 image. Coordinates are for 1950.0. Confidence
levels start at 3$\sigma$ with 1.5$\sigma$ steps. \label{Figure1}}
\end{figure}

\clearpage

\begin{figure}

\centerline{\psfig{figure=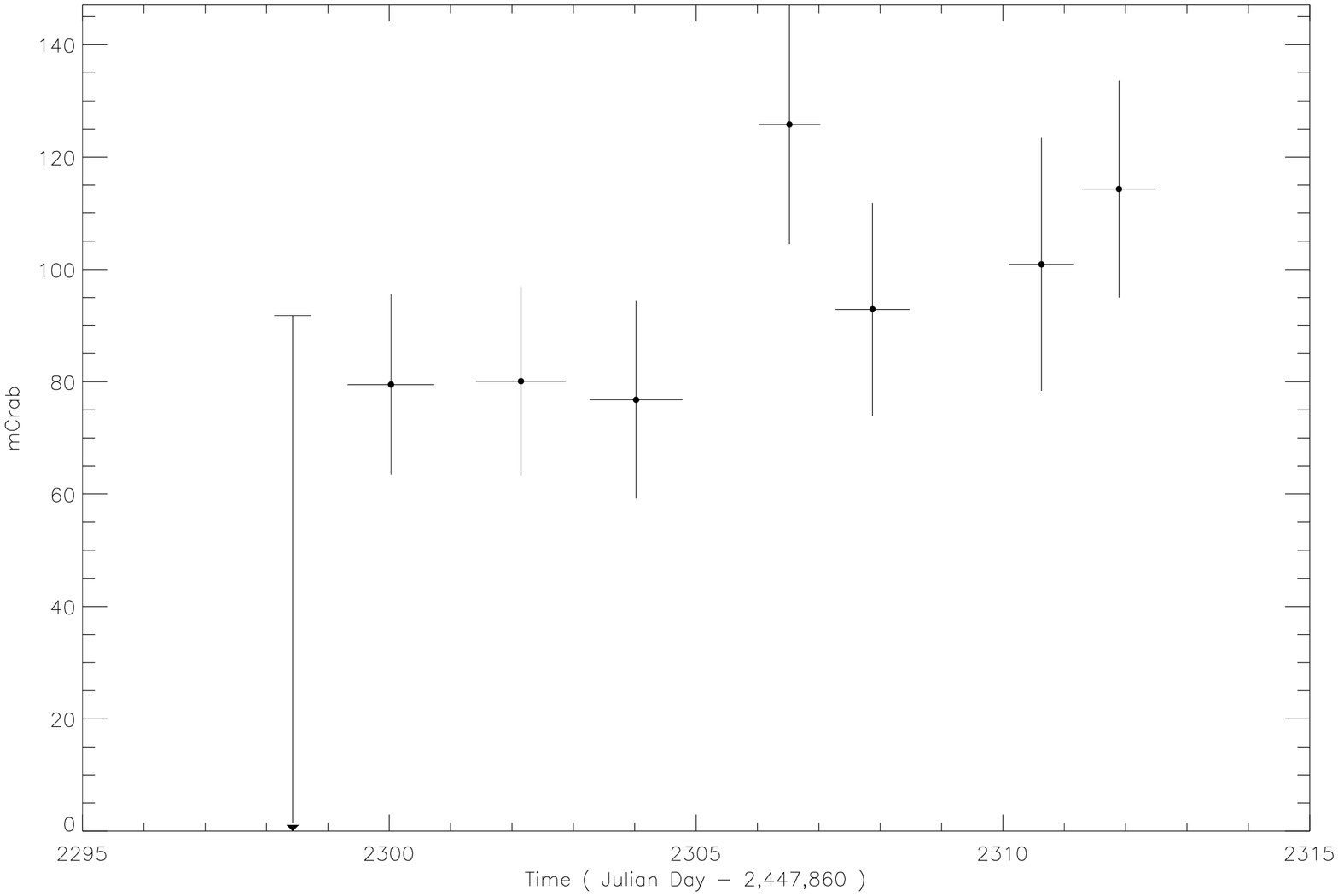,height=60mm,width=80mm,angle=90}
            \psfig{figure=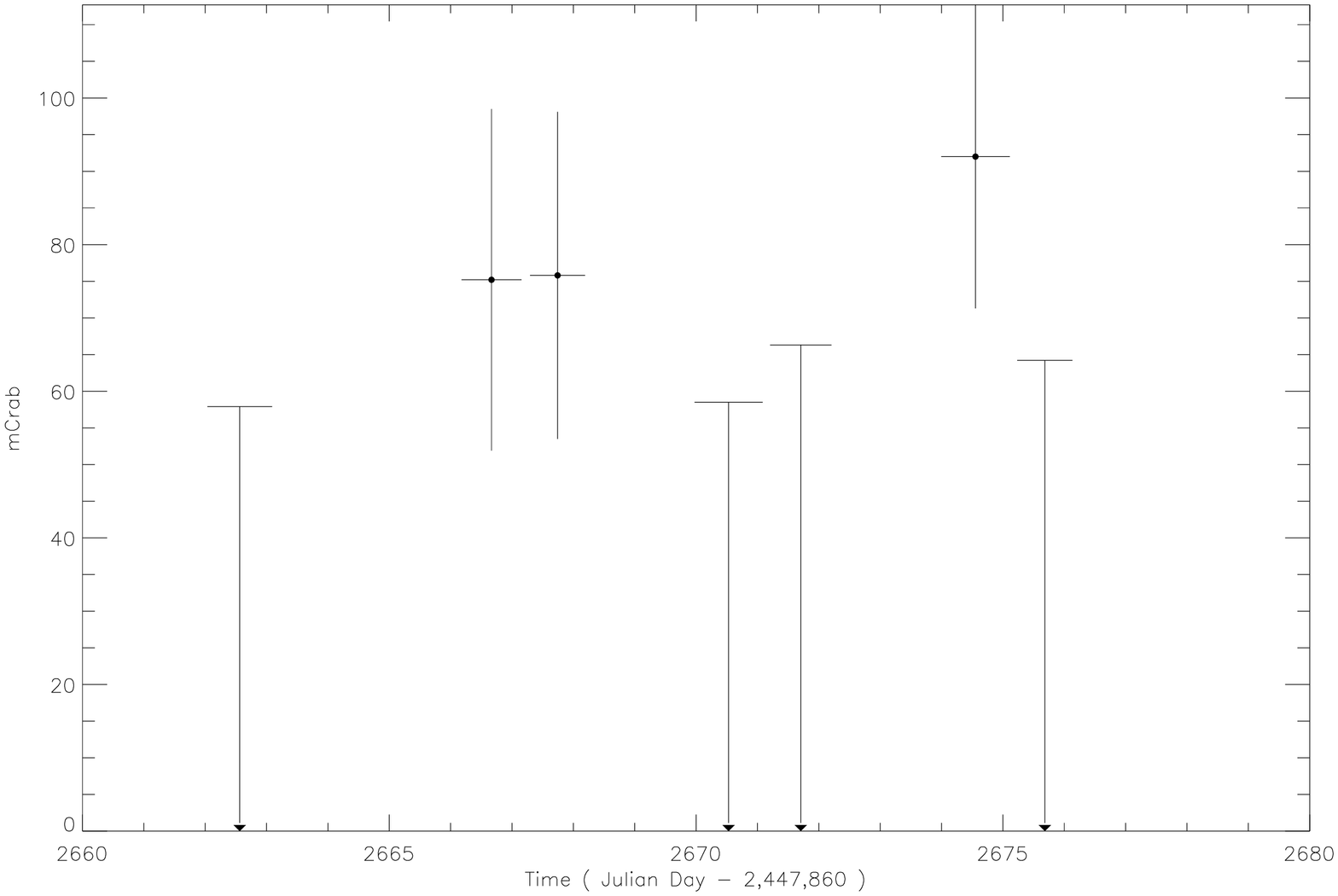,height=60mm,width=80mm,angle=90}}

\caption{Light curve of GRO J1744-28 in the 35-75 keV energy band for the March 1996
and March 1997 campaigns, as observed with SIGMA. The horizontal axis corresponds to the
date since the begining of the SIGMA observations on 1 Dec. 1989. \label{Figure2}}
\end{figure}

\clearpage

\begin{figure}

\centerline{\psfig{figure=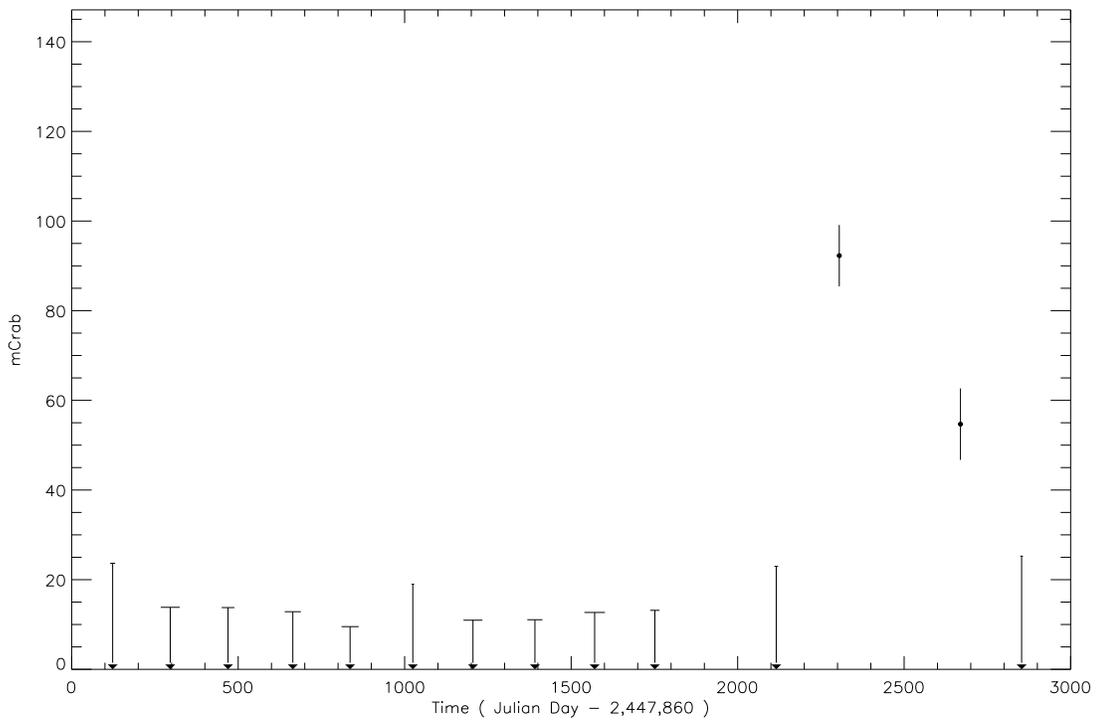,height=100mm,width=150mm,angle=90}}

\caption{Light curve of GRO J1744-28 in the 35-75 keV energy band for all the
SIGMA observing campaigns in the direction of the GC region, begining on 1 Dec. 1989. \label{Figure3}}
\end{figure}

\clearpage

\begin{figure}

\centerline{\psfig{figure=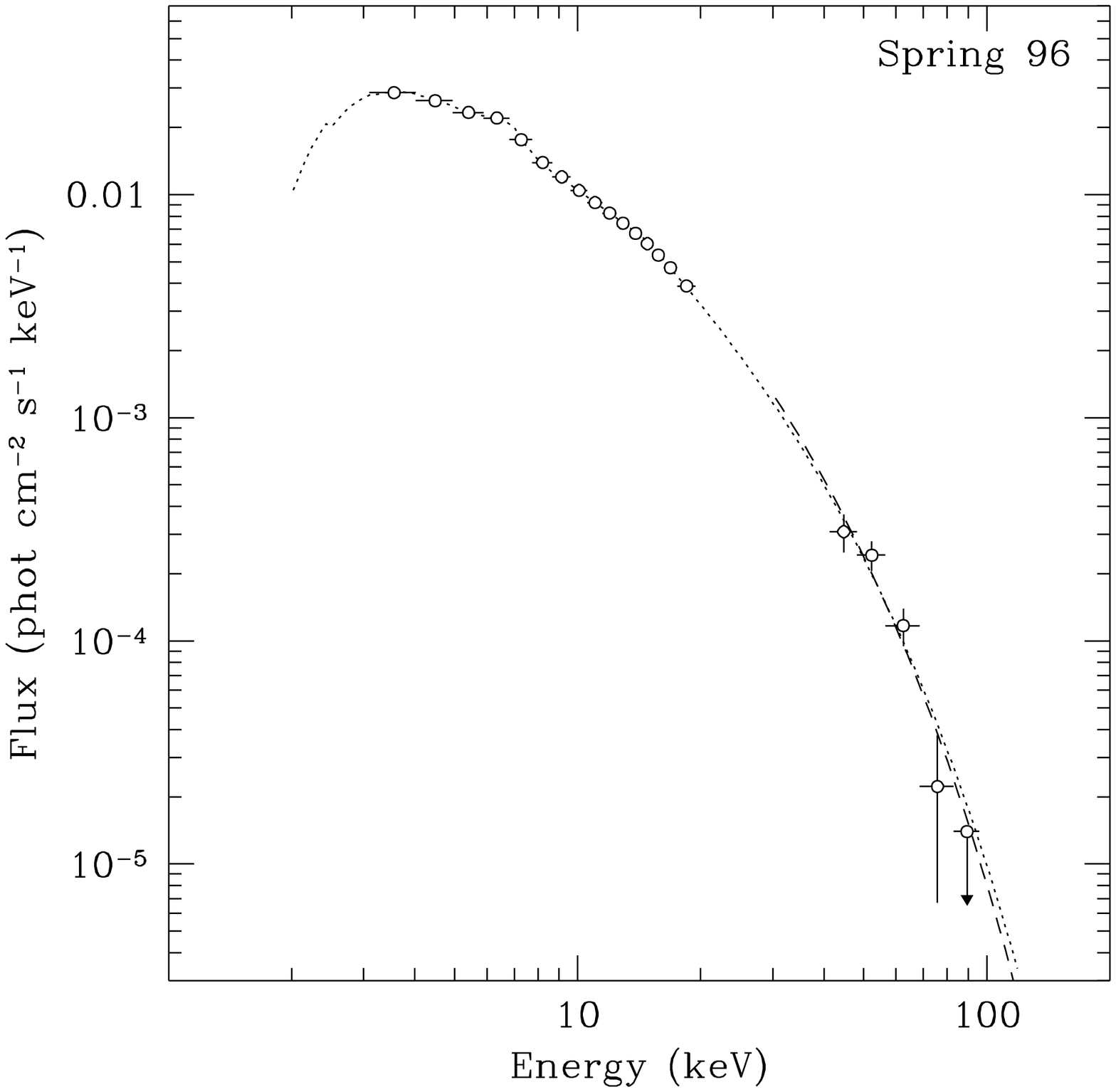 ,height=90mm,width=80mm}
            \psfig{figure=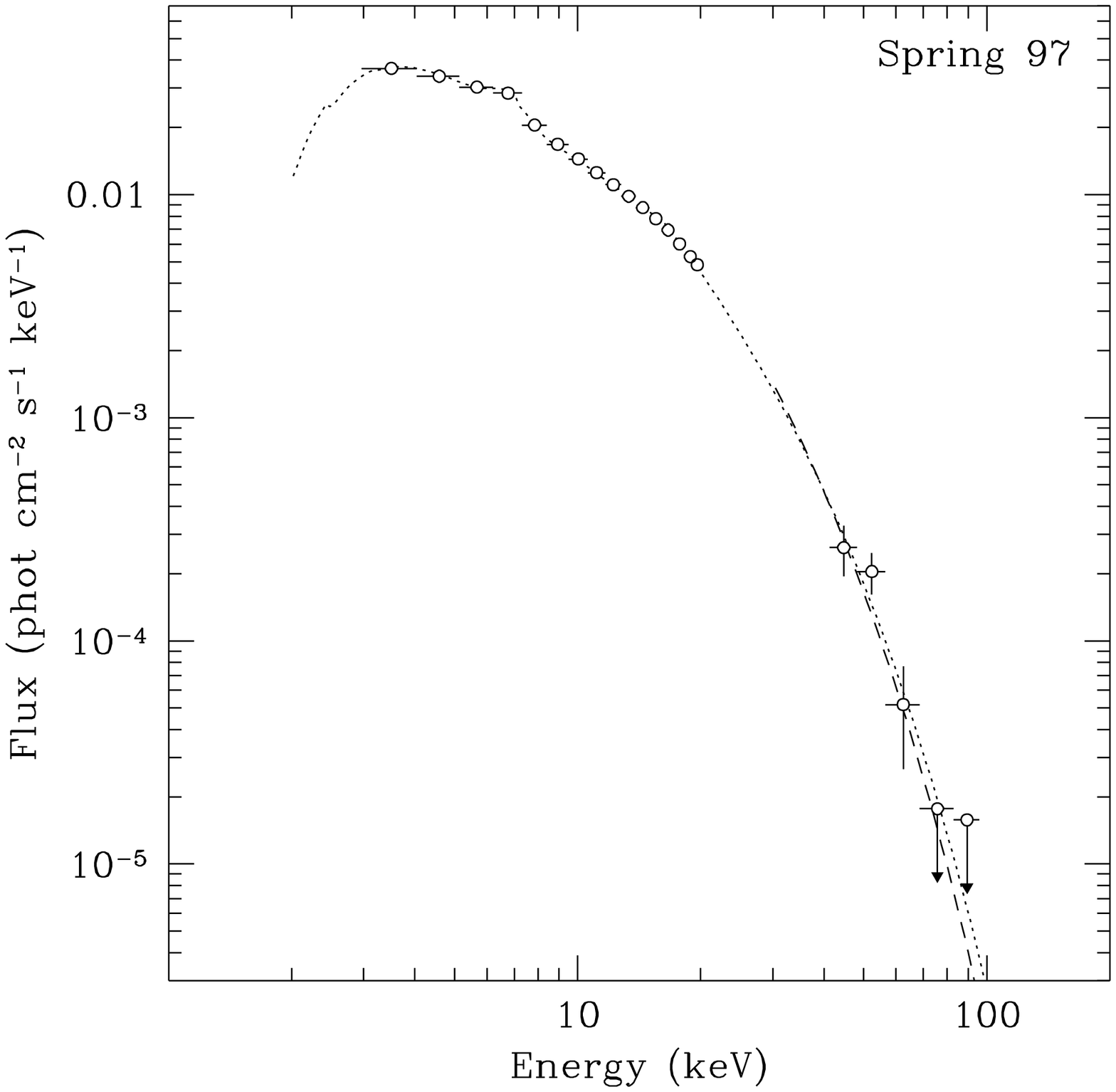 ,height=90mm,width=80mm}}

\caption{Combined RXTE/PCA and GRANAT/SIGMA spectra of GRO J1744-28 obtained in March 1996
and March 1997.
The PCA data points were renormalized to the SIGMA flux. The dotted line represents the best
fit model for the spectra (see text), with power law photon index $\alpha$$\sim$1.4 and
high energy cutoff E$\sim$15 keV in the first case and E$\sim$16 keV in the second. \label{Figure4}}
\end{figure}

\clearpage

\begin{table}
\caption{SIGMA March 1996 and March 1997 observation log of the Galactic Center 
region containing the coordinates of GRO J1744-28.}

\label{Table1}
\[
\begin{array}{cccccc}
\hline\hline
\noalign{\smallskip}
  ${\rm Session}$ & ${\rm Date}$ & ${\rm Exposure}$
& ${\rm 35-75 keV flux}$ & ${\rm 75-150 keV flux}$ &\\
& ${\rm (UT)}$ & ${\rm (hours)}$
& ${\rm (mCrab)}$ & ${\rm (mCrab)}$ &\\

\noalign{\smallskip}
\hline
\noalign{\smallskip}
& & & ${\rm March 1996 Campaign}$ & &\\
\noalign{\smallskip}
\hline
\noalign{\smallskip}
894 & 15.63-16.68 & 25.25 & 80 \pm 26 & <34 &\\
895 & 16.82-18.23 & 33.84 & 77 \pm 15 & <23 &\\
896 & 18.92-20.38 & 35.06 & 62 \pm 15 & <25 &\\
897 & 20.77-22.48 & 41.14 & 75 \pm 16 & <21 &\\
898 & 23.52-24.52 & 24.03 & 106 \pm 18 & 57 \pm 28 &\\
899 & 24.77-26.48 & 41.14 & 86 \pm 16 & <24 &\\
900 & 27.60-28.66 & 25.43 & 96 \pm 20 & <30 &\\
901 & 28.79-30.50 & 41.14 & 96 \pm 17 & <28 &\\
TOTAL & & 267.03 & 85 \pm 6.4 & <28 &\\
\noalign{\smallskip}
\hline
\noalign{\smallskip}
& & & ${\rm March 1997 Campaign}$ & &\\
\noalign{\smallskip}
\hline
\noalign{\smallskip}
935 & 14.54-15.59 & 25.30 & 34 \pm 17 & <30 &\\
937 & 18.68-19.65 & 23.45 & 75 \pm 22 & <29 &\\
938 & 19.79-20.69 & 21.49 & 69 \pm 21 & <33 &\\
939 & 22.47-23.58 & 26.66 & 43 \pm 17 & <28 &\\
940 & 23.71-24.71 & 24.00 & 22 \pm 21 & <28 &\\
941 & 26.49-27.61 & 26.81 & 83 \pm 18 & 46 \pm 26 &\\
942 & 27.73-28.63 & 21.58 & 34 \pm 20 & 58 \pm 31 &\\
TOTAL & & 148.53 & 52 \pm 7.4 & <33 &\\
\noalign{\smallskip}
\hline
\end{array} 
\]
\noindent 1 mCrab corresponds to $\sim$1.08$\times$10$^{-4}$ and $\sim$4.4$\times$10$^{-5}$ 
photons cm$^{-2}$ s$^{-1}$ ($\sim$8.6$\times$10$^{-12}$ and $\sim$7.3$\times$10$^{-12}$ 
ergs cm$^{-2}$ s$^{-1}$) for the 35-75 keV and 75-150 keV bands respectively.
\end{table}

\end{document}